\documentclass[letter,traditabstract, longauth]{aa}
\usepackage{graphicx,txfonts}
\begin{document}
\title{A {\it Herschel}\thanks{{\it Herschel} is an ESA space observatory 
with science instruments provided
by European-led Principal Investigator consortia and with 
important participation from NASA.}
study of the properties of starless cores
in the Polaris Flare dark cloud region using PACS and SPIRE}
\titlerunning{Starless cores in the Polaris Flare region}
\author{D. Ward-Thompson\inst{1} 
\and J. M. Kirk\inst{1}
\and P.~Andr\'e\inst{2}
\and P.~Saraceno\inst{3}
\and P.~Didelon\inst{2}
\and V.~K\"onyves\inst{2}
\and N.~Schneider\inst{2}
\and A.~Abergel\inst{4}
\and J.-P.~Baluteau\inst{5}
\and J.-Ph.~Bernard\inst{6}
\and S.~Bontemps\inst{2}
\and L.~Cambr\'esy\inst{7}
\and P.~Cox\inst{8}
\and J.~Di~Francesco\inst{9}
\and A.~M.~Di~Giorgio{3}
\and M.~Griffin\inst{1}
\and P.~Hargrave\inst{1}
\and M.~Huang\inst{10}
\and J.~Z.~Li\inst{10}
\and P.~Martin\inst{11}
\and A.~Men'shchikov\inst{2}
\and V.~Minier\inst{2}
\and S.~Molinari\inst{3}
\and F.~Motte\inst{2}
\and G.~Olofsson\inst{12}
\and S.~Pezzuto\inst{11}
\and D.~Russeil\inst{6}
\and M.~Sauvage\inst{2}
\and B.~Sibthorpe\inst{13}
\and L.~Spinoglio\inst{3}
\and L.~Testi\inst{14}
\and G.~White\inst{15}
\and C.~Wilson\inst{16}
\and A.~Woodcraft\inst{13}
\and A.~Zavagno\inst{5}}
\institute{School of Physics and Astronomy, Cardiff University, 
 Queens Buildings, The Parade, Cardiff, CF243AA, UK  %1 
\and Laboratoire AIM, CEA/DSM--CNRS--Universit\'e Paris Diderot, IRFU/ 
 Service d'Astrophysique, C.E. Saclay, Orme des Merisiers,
 91191 Gif-sur-Yvette, France                   %2
\and INAF-IFSI, Fosso del Cavaliere 100, 00133 Roma, Italy    %3
\and IAS, Universit\'e Paris-Sud, Bat 121, 91405 Orsay, France  %4
\and LAM/OAMP, Universit\'e de Provence, 13388 Marseille, France  %5
\and CESR, 9 Avenue du Colonel Roche, BP 4346, 31029 Toulouse, France %6
\and CDS, Observatoire de Strasbourg, 11 rue de l'Universit{\'e}, 
67000 Strasbourg, France                    %7
\and IRAM, 300 rue de la Piscine, Domaine Universitaire, 38406 Saint  
Martin d'H\'eres, France                     %8
\and Herzberg Institute of Astrophysics, Department of Physics and Astronomy, 
University of Victoria, Victoria, Canada       %9
\and NAOC, Chinese Academy of Sciences,
A20 Datun Road, Chaoyang District, Beijing 100012, China %10
\and CITA, University of Toronto, 60 St George Street, Toronto, Ontario, 
M5S 3H8, Canada              %11
\and Department of Astronomy, Stockholm University, AlbaNova University 
Center, SE-10691 Stockholm, Sweden   %12
\and UKATC, Royal Observatory, Blackford Hill, Edinburgh, EH93HJ, UK %13
\and INAF, Largo Enrico Fermi 5, I-50125  Firenze, Italy  %14
\and RAL, Chilton, Didcot, OX110NL, UK, and Open University, 
 Milton Keynes MK76AA, UK %15
\and Dept. of Physics \& Astronomy, McMaster University, Hamilton,  
Ontario, L8S 4M1, Canada        %16
}

\date{Received 07/05/2010. Accepted 10/05/2010. In original form 31/03/2010.}

\abstract
{The Polaris Flare cloud region contains a great deal 
of extended emission. It is at high declination and high Galactic 
latitude. It was previously seen strongly in IRAS Cirrus emission 
at 100 microns. We have detected it with both PACS and SPIRE on
{\it Herschel}. We see filamentary and low-level structure.
We identify the five densest cores within this
structure. We present the results of a temperature, mass and density 
analysis of these cores. We compare their observed masses to their virial
masses, and see that in all cases the observed masses lie 
close to the lower end 
of the range of estimated virial masses.
Therefore, we cannot say whether they are gravitationally bound 
prestellar cores. Nevertheless, these are the best candidates to be potential
prestellar cores in the Polaris cloud region.}

\maketitle

\section{Introduction}

In this paper we present observations, performed with the ESA 
{\it Herschel} Space
Observatory (Pilbratt et al 2010), of the Polaris Flare region.
In particular we use the large collecting area
and powerful science payload of {\it Herschel}
to perform imaging photometry using the PACS (Poglitsch et al 2010) and SPIRE
(Griffin et al 2010) instruments. These observations were carried out as 
part of the guaranteed-time key programme to map most of the Gould Belt 
star-forming regions with {\it Herschel} (Andr\'e et al 2010).
The Polaris Flare was first detected in HI as a spur of gas that appears
to rise more than 30$^\circ$ out of the Galactic Plane.
This region is an area rich in IRAS cirrus emission
(e.g. Low et al 1984), and is sometimes
known as the Polaris Cirrus Cloud. 
It was mapped in CO by Heithausen \& Thaddeus (1990). On
the large scale this cloud appears to merge with the Cepheus Flare cloud
(e.g. Kirk et al 2009), and both clouds extend to high Galactic latitude. 

\begin{figure*}
\centering{\includegraphics[width=\textwidth]{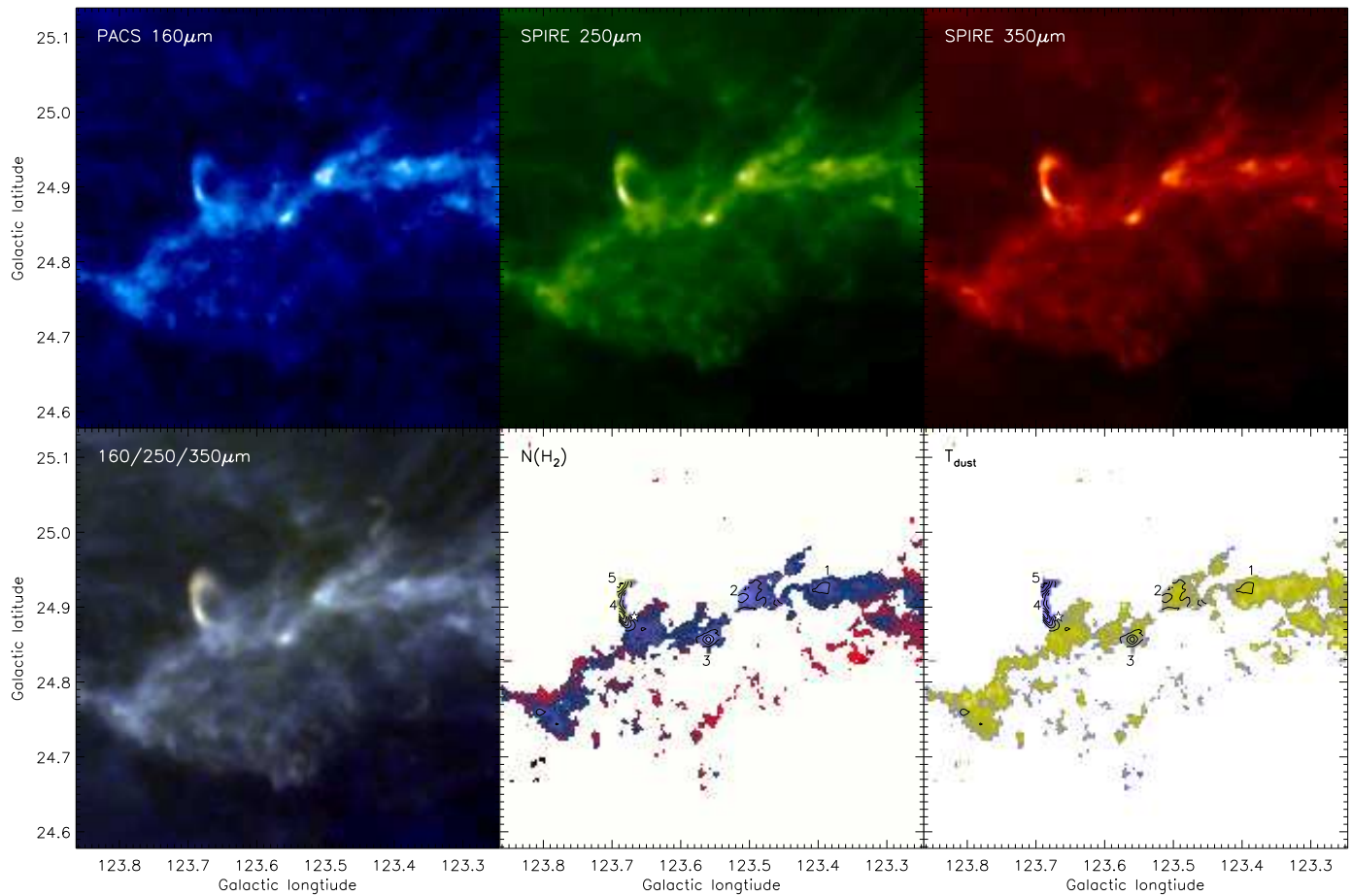}}
\caption{\label{5bands}The densest part of the Polaris Flare region at 
some of the observed wavebands. Upper row: 160$\mu$m from PACS, and 
250$\mu$m and 350$\mu$m from SPIRE. Lower row: False-colour image 
(where 160$\mu$m is shown in blue, 250$\mu$m is shown in green, and 
350$\mu$m is shown in red), column density map (where red is $<$4, 
blue is 4--8, and yellow is $>$8 $\times$ 10$^{21}$ cm$^{-2}$), and 
colour temperature map (where blue is 10--11~K and yellow is 12--13~K). 
The contour levels on the column density map start  at 4 $\times$ 10$^{21}$ 
cm$^{-2}$, and the interval between successive contours is 1.5 $\times$ 
10$^{21}$ cm$^{-2}$. The same contours are repeated on the temperature map 
for ease of location. Five sources are seen above a column density of 4 
$\times$ 10$^{21}$ cm$^{-2}$. These are labelled cores 1--5 (in order of 
increasing R.A.) on the last two panels and are discussed in the text. 
The loop (loop 1) discussed in the text (containing cores 4 \& 5) is 
clearly visible in all images. The reddest features on the false-colour 
image are the coldest, and the loop shows up clearly as redder than the 
surroundings. Likewise in the temperature map, the loop shows up as blue, 
indicating that it is the coldest feature on the map. The position of the 
IRAS source (IRAS~01432+8725) is marked with a star on the last two panels 
(adjacent to core 4).}
\end{figure*}

One of the denser regions in the cloud is known as molecular cloud 
123.5+24.9, or MCLD~123.5+24.9 (e.g. Bensch et al 2003) -- hereafter 
MCLD~123 -- at a 
distance of 150~pc (Bensch et al 2003). It shows strong extended IRAS 
100-$\mu$m emission and is generally believed to be gravitationally 
unbound with a mass of $\sim$18--32~M$_\odot$ (Grossmann et al 1990; 
Bensch et al 2003). A CO study by Falgarone et al (1998) revealed a 
curved filament in MCLD~123 in $^{13}$CO and C$^{18}$O -- both in the 
J=2--1 transition. This filament is also apparent in some narrow velocity 
channels in the same transition of $^{12}$CO (Falgarone et al 1998).

There is one IRAS source in the region, IRAS~01432+8725. This is listed 
in the IRAS catalogue as having a flux density at 100~$\mu$m of 2.88~Jy, 
but only upper limits at the other IRAS wavebands. There is also one 
{\it Spitzer}
source that was only detected at a wavelength of 24~$\mu$m at coordinates 
R.A. (2000) $=$ 01$^{\rm h}$ 58$^{\rm m}$ 27.5$^{\rm s}$, Dec. (2000) 
$=$ $+$87\degr\ 40\arcmin\ 07\arcsec\ . It has a peak flux density at 
24~$\mu$m of 1.3~mJy/beam, where the {\it Spitzer} beam at this wavelength is 
7~arcsec. This detection lies in a {\it Spitzer} calibration field in an 
unpublished archival dataset (AOR~33136386). 

\section{Observations}

The SPIRE/PACS parallel-mode Science Demonstration Observations of the 
Polaris cloud were performed on 2009 October 23 (Operation Day 162) 
at wavelengths of 70~$\mu$m and 160~$\mu$m with PACS, and at 250~$\mu$m, 
350~$\mu$m and 500~$\mu$m with SPIRE. The 70- and 160-$\mu$m 
$\sim$6\,deg$^2$ scan map was taken with 60~arcsec/sec scanning speed. 
The field was observed twice with both instruments by performing cross-linked
scans in two nearly orthogonal scan directions. The combination of nominal
and orthogonal coverages reduced the effects of 1/f noise and better 
preserved spatial resolution. The SPIRE data were reduced using HIPE 
version 2.0 and the pipeline scripts delivered with this version. These 
scripts were modified, e.g. observations that were taken during the 
turnaround of the satellite were included. A median baseline (HIPE default)
was applied to the maps and the `naive mapper' was used for map making.

The PACS data were reduced with HIPE 3.0.455 provided by the {\it Herschel} 
Science Center (HSC). We used file version 1 flat-fielding and responsivity 
in the calibration tree, instead of the built-in version 3. Therefore the 
error in the final reduced flux scale was corrected manually with the 
corresponding correction values in the PACS wavelengths. Standard steps 
of the default pipeline were applied for data reduction starting from 
(level 0) raw data. Multi-resolution median transform (MMT) deglitching 
and second order deglitching were also applied. Baselines were subtracted 
from the level 1 data by high-pass filtering with a $\sim$1$^\circ$ filter 
width, avoiding obvious sources, whilst the full leg length was 2.5$^\circ$
in the parallel mode.

\begin{table*}
\begin{center}
\caption{The physical properties of the cores. The Galactic latitude and 
longitude, as well as Right Ascension and Declination, are listed, 
along with the assumed distance. The radius
of each core was measured by taking the column density map in Fig.~1 and 
measuring the equivalent radius of the contour that encircled a column 
density of 4 $\times$ 10$^{21}$ cm$^{-2}$, and this radius is given in pc. 
The full-width at half maximum (FWHM) is the geometric mean FWHM measured
on the peak of each source.
The integrated flux density within this contour at each of the {\it Herschel} 
wavelengths is listed in Jy. The absolute uncertainty in the flux densities 
is $\pm$15\%. As described in the text, a temperature, peak column density 
and mass were derived, and these are also listed. The uncertainty in the 
masses could be as high as a factor of 2. A mean volume density is given, 
assuming each core is spherical, within the given radius. Finally, a virial 
mass for each core is estimated using CO and HCO$+$ linewidths.}
\begin{tabular}{lrrrrr} \hline
Parameter & Core 1 & Core 2 & Core 3 & Core 4 & Core 5 \\ \hline 
Galactic longitude (2000) & 123.388 & 123.511 & 
123.559 & 123.687 & 123.690 \\
Galactic latitude (2000) & $+$24.928 & $+$24.915 & 
$+$24.856 & $+$24.894 & $+$24.931\\
Right Ascension (2000) & 
01$^{\rm h}$ 34$^{\rm m}$ 01.9$^{\rm s}$ & 
01$^{\rm h}$ 44$^{\rm m}$ 51.6$^{\rm s}$ & 
01$^{\rm h}$ 47$^{\rm m}$ 40.8$^{\rm s}$ & 
01$^{\rm h}$ 59$^{\rm m}$ 42.7$^{\rm s}$ & 
02$^{\rm h}$ 00$^{\rm m}$ 58.7$^{\rm s}$ \\
Declination (2000) & 
$+$87$^{\circ}$ 45$^{\prime}$ 42$^{\prime\prime}$ & 
$+$87$^{\circ}$ 43$^{\prime}$ 35$^{\prime\prime}$ & 
$+$87$^{\circ}$ 39$^{\prime}$ 33$^{\prime\prime}$ & 
$+$87$^{\circ}$ 39$^{\prime}$ 53$^{\prime\prime}$ & 
$+$87$^{\circ}$ 41$^{\prime}$ 58$^{\prime\prime}$ \\
Distance (pc) & 150 & 150 & 150 & 150 & 150 \\
Radius (pc) & 0.023 & 0.035 & 0.032 & 0.035 & 0.034 \\
FWHM (pc) & 0.023 & 0.039 & 0.027 & 0.042 & 0.038 \\
F$^{int}_{70\mu m}$  (Jy) & $<$0.18 & $<$0.18 & 
$<$0.18 & $<$0.18 & $<$0.18 \\
F$^{int}_{160\mu m}$ (Jy) & 4.26 $\pm$ 0.07 &  10.19 $\pm$ 0.07 &    
7.56 $\pm$ 0.07 & 7.63 $\pm$ 0.07 & 5.44 $\pm$ 0.07 \\
F$^{int}_{250\mu m}$ (Jy) & 6.74 $\pm$ 0.04 & 16.77 $\pm$ 0.04 &   
13.10 $\pm$ 0.04 & 15.35 $\pm$ 0.04 & 13.26 $\pm$ 0.04 \\
F$^{int}_{350\mu m}$ (Jy) & 3.61 $\pm$ 0.02 &  8.98 $\pm$ 0.02 &    
7.27 $\pm$ 0.02 &  9.05 $\pm$ 0.02 &  8.50 $\pm$ 0.02 \\
F$^{int}_{500\mu m}$ (Jy) & 1.72 $\pm$ 0.02 &  4.33 $\pm$ 0.02 &   
3.50 $\pm$ 0.02 &  4.53 $\pm$ 0.02 &  4.40 $\pm$ 0.02 \\
Temperature (K) & 12 $\pm$ 1 & 12 $\pm$ 1 & 12 $\pm$ 1 & 11 $\pm$ 1 &
10 $\pm$ 1 \\
N(H$_2$)$_{peak}$ ($\times$10$^{21}$ cm$^{-2}$) & 6 $\pm$ 3 & 7 $\pm$ 3 &
9 $\pm$ 4 & 13 $\pm$ 5 & 13 $\pm$ 5 \\
Mass (M$_\odot$) & 0.1 & 0.3 & 0.3 & 0.4 & 0.5 \\ 
n(H$_2$)$_{mean}$ (cm$^{-3}$) & $\sim$5$\times$10$^{4}$ & 
$\sim$4$\times$10$^{4}$ &
$\sim$4$\times$10$^{4}$ & $\sim$5$\times$10$^{4}$ & $\sim$7$\times$10$^{4}$\\
M$_{vir}$ (M$_\odot$) & $\sim$0.3--0.5 & $\sim$1.0--1.5 & $\sim$1.0--1.5 & 
$\sim$1.0--1.5 & $\sim$1.0--1.5 \\  \hline
\end{tabular}
\end{center}
\label{coreprop}
\end{table*}

The PACS data of this field include transients of unknown origin after 
each calibration block, which seriously affected the ensuing frames. We 
processed these observations using data-masking and a narrower high-pass 
filter width than the image size in order to mitigate the calibration block 
artifacts. In this process, we may have removed spatial scales larger than 
the filter widths. The final PACS maps were created using the HIPE `MADmap' 
mapping method projected to the
3.2 and 6.4 arcsec/pixel size for 70 and 160\,$\mu$m data, respectively.

\section{Results}

The Polaris Flare dark cloud region was observed at five wavelengths -- 70, 
160, 250, 350 and 500$\mu$m. Figure~1 shows some of the main results. Only 
the densest part of the mapped region is shown. The area shown is just over 
half a degree square. The upper row of Fig.~1 shows the data from three of 
the wavebands: 160$\mu$m from PACS; and 250$\mu$m and 350$\mu$m from SPIRE. 
The data have been smoothed to a common resolution of 24~arcsec, the 
approximate resolution of the 350-$\mu$m data. The images have also been 
re-gridded onto 10 $\times$ 10 arcsec pixels.

The lower row of Fig.~1 shows some images derived from the raw data: a 
false-colour image; a column density map;
and a colour temperature map.
The contours on the column density map are  at 4, 5.5, and 
7 $\times$ 10$^{21}$ cm$^{-2}$. These are repeated on the temperature map 
to assist in source location. The Polaris cloud is clearly seen, and the 
raw data show a complex structure that is broadly similar at all wavebands. 
There are a number of filamentary structures seen in the data, with a few 
brighter cores embedded in the cloud.

There is a filamentary loop seen in all images that is centred roughly at
Galactic coordinates l = 123.67, b = $+$24.89 -- 
R.A.(2000) = 01$^{\rm h}$ 58$^{\rm m}$, Dec.(2000) = 87$^\circ$ 40$^\prime$.
We here label this feature loop~1. This is the same curved filament as was 
seen by Falgarone et al (1998) in $^{13}$CO. They interpreted this as an 
edge of a cloud core. However, in the continuum we see it is clearly a 
loop with no filled centre. It was also detected in various transitions
by Grossman \& Heithausen (1992).

There is also a filament with an apparent bifurcation at roughly Galactic 
coordinates l = 123.48, b = $+$24.90 -- R.A.(2000) = 01$^{\rm h}$ 
42$^{\rm m}$, Dec.(2000) = 87$^\circ$ 43$^\prime$. 
A bright core region is seen at the 
head of this bifurcation, which may be broken up into three components 
in the 160-$\mu$m data.
The mean off-source pixel-by-pixel 1-$\sigma$ variation
on the N(H$_2$) map varies from 1.2 to 1.5 $\times$ 10$^{21}$ cm$^{-2}$.
Hence, we adopt a value of 4 $\times$ 10$^{21}$ cm$^{-2}$ for the
3-$\sigma$ contour.

Five sources are seen in the column density map above a column density of 
4 $\times$ 10$^{21}$ cm$^{-2}$. We here label these cores~1--5 in order 
of increasing Galactic longitude -- see lower right panels of Fig.~1. 
We list the core positions and their assumed distances in Table~1. The 
core mentioned above at the bifurcated filament is core~2, and loop~1 
contains cores 4 and 5.

The IRAS source IRAS~01432+8725 lies an arcminute to the west of core~4. 
We believe this offset is sufficient that the two sources are different 
(the IRAS FWHM at 100~$\mu$m is 44~arcsec). Therefore, none of the cores 
is associated with an infrared source, and so these are all candidate 
starless cores (Myers et al., 1987). The IRAS source is coincident with 
the centre of the loop, and may in fact be 
loop~1 itself, as IRAS point sources that only show up at 100~$\mu$m 
have often in the past been shown to be simply bits of cirrus. 
The {\it Spitzer}
source may be foreground, as it is only seen at the shortest wavelengths.

The reddest features on the false-colour image are the coldest, and 
loop~1 shows up clearly as redder than the surroundings. Likewise in 
the temperature map, the loop shows up as blue, indicating that it is 
the coldest feature on the map. Cores 4 \& 5 appear to be the densest
features on the map, with peak column densities in excess of 
10$^{22}$ cm$^{-2}$. The column density contour of 4 $\times$ 10$^{21}$ 
was selected as the core boundary in each case. The radial sizes of the 
cores were estimated from the images as the equivalent radius of a circle 
with an area equal to that contained by the core boundary. The derived 
equivalent radii are listed in Table~1. Flux densities were measured 
within the core boundary contour in each case, and these are also 
listed in Table~1.

\begin{figure*}
\centering{\includegraphics[width=\textwidth]{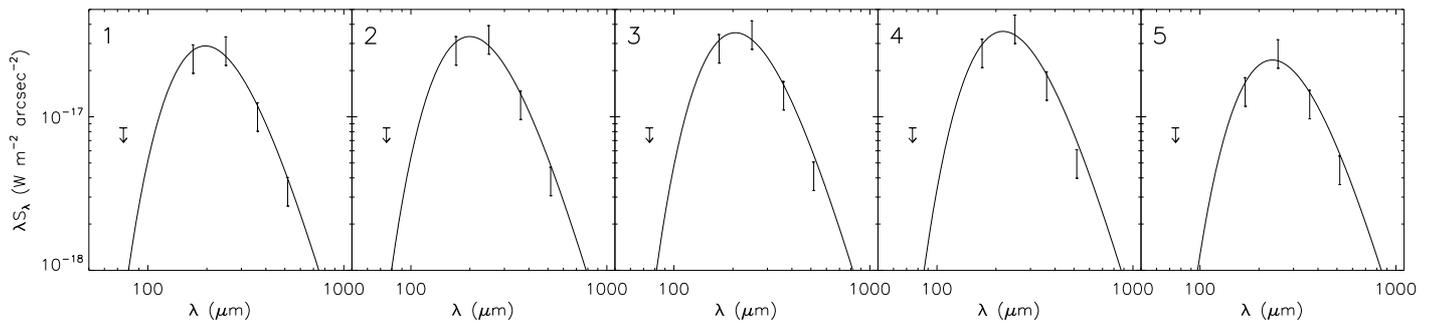}}
\caption{\label{sed} Spectral energy distributions of cores 1 to 5. 
The peak flux density in a single 10 $\times$ 10 arcsec pixel was 
measured. This is shown on a log-log plot of $\lambda S_\lambda$ 
versus $\lambda$. The data are shown with 15\% uncertainty error-bars. 
The upper limits at 70~$\mu$m are shown as arrows. The solid lines are 
grey-body fits of the form described in the text. The temperatures of 
the fits are listed in Table~1.}
\end{figure*}

\section{Core properties}

The core properties were estimated from the maps of column density and 
temperature. The flux densities of the pixels coincident with the column 
density peaks of each core are plotted on the spectral energy distributions
(SEDs) shown in Fig.~\ref{sed}. Modified blackbody curves were fitted to 
the flux densities, and these are also shown in Fig.~\ref{sed}. These are 
the same fits that were used, pixel-by-pixel, to construct the column 
density and temperature maps shown in Fig. \ref{5bands}. The form of 
the fit (c.f. Hildebrand 1983) that was used in each case is

\begin{equation}
F_\nu = \Omega B_\nu(T) ( m_{\rm H} \mu {\rm N}({\rm H}_2) \kappa_\nu ) \, ,
\end{equation}

\noindent
where $F_\nu$ is the flux density at frequency $\nu$, $\Omega$ is the 
solid angle of each pixel, $B_\nu(T)$ is the blackbody function at 
temperature $T$, $m_{\rm H}\mu$ is the mean particle mass ($m_{\rm H}$ 
is the mass of a hydrogen atom and $\mu$ was taken to be 2.86, assuming 
the gas is $\sim$70\% H$_2$ by mass), 
N(H$_2$) is the column number density of molecular hydrogen, 
and $\kappa_\nu$ is the dust mass opacity.

We used the pixel-by-pixel SED fits to calculate the column densities, 
and hence the core masses.
The value of $\kappa_\nu$ that should be used has been the subject of 
much controversy. Here we adopt the dust opacity recommended by Henning 
et al (1995) and Preibisch et al (1993) for clouds of intermediate 
density -- n(H$_2$) $\leq$ 10$^5$ cm$^{-3}$ -- and we assume a standard 
gas to dust mass ratio of 100. This is a similar parameterization of the 
dust opacity to that used by Beckwith et al (1990), namely that

\begin{equation}
\kappa_\nu = 0.1 \textrm{cm}^2\textrm{\,g}^{-1} 
\times (\nu/1000 {\rm Ghz})^{\beta},
\end{equation}

\noindent
where we have set the dust opacity index $\beta$ to be equal to 2. 

This is also consistent with the value used by Andr\'e et al (1993; 1996) 
and by Kirk et al (2005) for prestellar and starless cores. The peak 
column densities and the temperature at the peak are listed in Table~1. 
The mass of each core was calculated by integrating the column density map 
within the selected core boundary. This is also listed in Table~1. 
From these, the volume densities were calculated, 
assuming that the cores are spherical.
These, too, are listed in Table~1.

The temperatures of the SEDs are listed in Table~1. These are all quite low, 
with values of 10--12~K. This, and the lack of NIR emission from the cores, 
implies that the star formation process has yet to begin within these 
particular cores. Hence they are starless cores. This means that these 
cores should have no internal heating and should be heated solely by the 
external radiation field. This is similar to what is seen in other low-mass 
starless and prestellar cores (e.g. Ward-Thompson et al 2002; or for a 
review see Ward-Thompson et al 2007). 

Core 5 was observed in the submillimetre by Bernard et al (1999). Our 
results are consistent with their findings, allowing for the very different 
resolutions of the two sets of observations. Falgarone et al (2009) measured 
the mean CO linewidths in MCLD~123. For so-called `bright' regions (i.e. 
high column densities) they found a mean linewidth of 0.4~kms$^{-1}$ in 
this region. Heithausen et al. (2008) observed MCLD~123 in a number of 
transitions and found mean linewidths from 0.2 to 0.4~kms$^{-1}$.
Using this range of values we estimated a range of values for 
the virial masses of the five cores, and list these in Table~1. 

We note that all of the cores have masses that are below the 
virial masses that we have estimated. However, given the
uncertainties in the mass calculations, they could be
consistent with the lower limit of the range of
virial masses. Hence, we can only say that they may or may not be 
gravitationally bound, and
these 5 cores may be on the edge of possibly becoming prestellar cores.
Note that this is very different from the cores found in the Aquila 
region (Andr\'e et al, 2010), a large fraction of which are clearly
gravitationally bound -- c.f. Fig.~4 of Andre et al (2010).  
Nevertheless, bound or unbound,
the 5 cores we have selected are the closest to being gravitationally
bound of any of the starless cores in Polaris.
 
\section{Conclusions}

We have presented {\it Herschel} data of the Polaris Flare dark cloud region, 
and in particular the region MCLD~123. We found a great deal of extended 
emission at wavelengths from 70 to 500~$\mu$m with both PACS and SPIRE. 
We noted some filamentary and low-level structure. We identified the five 
densest cores within this structure. We carried out a temperature, mass and 
density analysis of the cores. We compared their observed masses to their 
virial masses, and found that the observed masses are on the lower limit of
the range of their estimated virial masses, and thus we cannot say for certain 
whether they are gravitationally bound.

\begin{acknowledgements} 
JMK acknowledges STFC for funding, while this work was carried out,
under the auspices of the Cardiff Astronomy Rolling Grant.
SPIRE was developed by a consortium of institutes led by Cardiff Univ. 
(UK) and including Univ. Lethbridge (Canada); NAOC (China); CEA, LAM 
(France); IFSI, Univ. Padua (Italy); IAC (Spain); Stockholm Observatory 
(Sweden); Imperial College London, RAL, UCL-MSSL, UKATC, Univ. Sussex (UK); 
Caltech, JPL, NHSC, Univ. Colorado (USA). This development has been 
supported by national funding agencies: CSA (Canada); NAOC (China); CEA, 
CNES, CNRS (France); ASI (Italy); MCINN (Spain); Stockholm Observatory 
(Sweden); STFC (UK); and NASA (USA). PACS was developed by a consortium 
of institutes led by MPE (Germany) and including UVIE (Austria); KUL, CSL, 
IMEC (Belgium); CEA, LAM (France); MPIA (Germany); IFSI, OAP/AOT, 
OAA/CAISMI, LENS, SISSA (Italy); IAC (Spain). This development has 
been supported by the funding agencies BMVIT (Austria), ESA-PRODEX 
(Belgium), CEA/CNES (France), DLR (Germany), ASI (Italy), and CICT/MCT (Spain).
\end{acknowledgements}

\end{document}